\def\aa{{ A\&A}}
\def\aas{{ A\&AS}}

\def\aj{{AJ}}

\def\al{$\alpha$}
\def\bet{$\beta$}
\def\amin{$^\prime$}
\def\annrev{{ARA\&A}} 

\def\apj{{ApJ}}

\def\apjs{{ApJS}}

\def\asec{$^{\prime\prime}$}

\def\cc{cm$^{-3}$}

\def\deg{$^{\circ}$}

\def\e#1{$\times$10$^{#1}$}
\def\etal{{et al. }}

\def\flux{ergs s$^{-1}$ cm$^{-2}$}

\def\kms{km\thinspaces$^{-1}$}
\def\lamb{$\lambda$}
\def\lum{ergs s$^{-1}$}
\def\micron{{$\mu$m}}
\def\mnras{{MNRAS}}

\def\nat{{Nature}}

\def\pasp{{PASP}}

\def\percm2{cm$^{-2}$}
\def\perhz{Hz$^{-1}$}

\def\lax    {${_<\atop^{\sim}}$ }
\def\gax    {${_>\atop^{\sim}}$ }

\def\kms    {~km~s$^{-1}$}
\def\refindent{\par\noindent\parskip=2pt\hangindent=3pc\hangafter=1 }

\def\>Z{$>$}
\def\<{$<$}

\def\simlt{\lower.5ex\hbox{$\; \buildrel < \over \sim \;$}}
\def\simgt{\lower.5ex\hbox{$\; \buildrel > \over \sim \;$}}
\def\sqr#1#2{{\vcenter{\hrule height.#2pt
      \hbox{\vrule width.#2pt height#1pt \kern#1pt
         \vrule width.#2pt}
      \hrule height.#2pt}}}

\def\oii{[\ion{O}{2}]}
\def\4363{[\ion{O}{3}]}
\def\heii{\ion{He}{2}}
\def\oiii{[\ion{O}{3}]}

\def\oi{[\ion{O}{1}]}
\def\siii{\ion{S}{3}}

\def\fex{[\ion{Fe}{10}]}
\def\nii{[\ion{N}{2}]}
\def\hei{\ion{He}{1}}
\def\sii{[\ion{S}{2}]}
\def\siii{[\ion{S}{3}]}
\def\hi{\ion{H}{1}}
\def\hii{\ion{H}{2}}

\def\farcs{\hbox{$.\!\!^{\prime\prime}$}}

\def\micron{\hbox{$\mu$m}}

\documentstyle[11pt,aaspp4]{article}
\received{}
\accepted{}

\slugcomment{To appear in {\it The Astrophysical Journal Supplement 
Series}}

\begin{document}

\title{A Search for ``Dwarf'' Seyfert Nuclei. III. Spectroscopic 
Parameters and Properties of the Host Galaxies}

\author{Luis C. Ho}
\affil{Department of Astronomy, University of California, Berkeley, CA 
94720-3411}

\and

\affil{Harvard-Smithsonian Center for Astrophysics, 60 Garden St., Cambridge, 
MA 02138\footnote{Present address.}}

\author{Alexei V. Filippenko}
\affil{Department of Astronomy, University of California, Berkeley, CA 
94720-3411}

\and
 
\author{Wallace L. W. Sargent}
\affil{Palomar Observatory, 105-24 Caltech, Pasadena, CA 91125}

\begin{abstract}
We have completed an optical spectroscopic survey of the nuclear regions 
($r$ \lax 200 pc) of a large 
sample of nearby galaxies.  Although the main objectives of the survey are 
to search for low-luminosity active galactic nuclei and to quantify their 
luminosity function, the database can be used for a variety of other purposes. 
This paper presents measurements of the spectroscopic parameters for the 
418 emission-line nuclei, along with a compilation of the global properties of
all 486 galaxies in the survey.  Stellar absorption generally poses a 
serious obstacle to obtaining accurate measurement of emission lines in 
nearby galactic nuclei.  We describe a procedure for removing the starlight 
from the observed spectra in an efficient and objective manner.  The main 
parameters of the emission lines (intensity ratios, fluxes, profile widths, 
equivalent widths) are measured and tabulated, as are several stellar 
absorption-line and continuum indices useful for studying the stellar 
population.  Using standard nebular diagnostics, we determine the probable 
ionization mechanisms of the emission-line objects.  The resulting spectral 
classifications provide extensive information on the demographics of 
emission-line nuclei in the local regions of the universe.  This new catalog 
contains over 200 objects showing spectroscopic evidence for recent star 
formation and an equally large number of active galactic nuclei, including 46 
which show broad H\al\ emission.  These samples will serve as the basis of 
future studies of nuclear activity in nearby galaxies.

\end{abstract}

 
\section{Introduction}

We have recently completed an extensive optical spectroscopic survey of the 
nuclei ($r$ \lax 200 pc) of nearby galaxies.  The main goals of the survey 
are to systematically search for and to investigate the physical properties of 
low-luminosity, or ``dwarf,'' active galactic nuclei (AGNs), but the rich 
database can be used for a variety of other applications.  In previous papers 
of this series, we have outlined the scientific objectives and observational 
parameters in greater detail (Filippenko \& Sargent 1985, hereafter Paper I; 
Ho, Filippenko, \& Sargent 1995, hereafter Paper II), and several companion 
papers present various aspects of the analysis (Ho, Filippenko, \& Sargent 
1997a, b, c, d; Ho \etal 1997e).  In brief, high-quality, 
long-slit, optical spectra of moderate resolution were obtained of the nucleus 
of almost every bright galaxy in the northern ($\delta\,>\, 0$\deg) sky.  
The chosen magnitude limit is $B_T$ = 12.5 mag, and the objects were selected 
from the Revised Shapley-Ames Catalog of Bright Galaxies (RSA; Sandage \& 
Tammann 1981). The spectra were acquired  using the Hale 5~m telescope at 
Palomar Observatory and cover the regions $\sim$4230--5110 \AA\ and 
$\sim$6210--6860 \AA, with spectral resolutions of approximately 4 \AA\ and 
2.5 \AA, respectively.  Paper II presents a complete spectral atlas of the 486 
galaxies contained in the Palomar survey.  

The observational distinction between low-luminosity and high-luminosity AGNs 
is largely arbitrary.  Our survey contains mostly nearby galaxies and is 
designed to select objects on the faint end of the luminosity function.  Thus, 
our study probes much lower luminosities than in other, more conventional 
AGN samples.  The Palomar AGNs have a median (extinction-corrected) H\al\ 
luminosity of only 2\e{39} \lum, and over 85\% of the sources lie below 
L(H\al) = 10$^{40}$ \lum\ (Ho \etal 1997a).  By contrast, the typical Seyfert 
nucleus in the widely-studied Markarian catalog emits roughly 10$^{41}$ \lum\ 
in H\al\ (estimated from the compilations of Mazzarella \& Balzano 1986 and 
Whittle 1992a).  Operationally, we will henceforth define ``low-luminosity'' 
or ``dwarf'' AGNs to be those with L(H\al) $\leq$ 10$^{40}$ \lum.

This paper focuses on the quantification of spectroscopic parameters from the 
Palomar survey.  The measurements of the emission lines are presented in \S\ 2, 
where we also discuss in depth our method of starlight subtraction.  Section 
3 describes our method of spectral classification and the results of 
applying it to our survey, while section 4 presents the measurements 
of several stellar absorption-line and continuum indices useful for studying 
the stellar population.  Finally, for easy reference and subsequent analysis, 
we summarize in \S\ 5 a number of global properties of the host galaxies.

\section{Emission Lines}

\subsection{Starlight and Continuum Subtraction}

The optical spectra of many emission-line nuclei, particularly those residing 
in host galaxies of early to intermediate Hubble types (Sbc and earlier), are 
contaminated heavily, and often dominated, by the stellar component.  As 
late-type giants dominate the integrated light of galaxy bulges, stellar 
absorption lines affect the strengths of most emission lines of interest.  
The magnitude of the effect obviously depends on the equivalent widths of the 
emission lines, but it is generally large in the centers of most ``normal'' 
galaxies, where the emission lines are quite weak.  Consequently, quantitative 
studies of most emission-line nuclei depend sensitively on the accuracy with 
which the stellar absorption lines can be removed.

We adopt a modified version of the technique of ``template subtraction'' to 
correct for starlight contamination.  The strategy is straightforward: 
subtract from the spectrum of interest a suitably scaled template spectrum 
(model) which best represents the continuum and absorption-line strength 
of the stellar component; the resulting end-product should then be a 
continuum-subtracted, pure emission-line spectrum.  The template model 
is constructed from a library of galaxy absorption-line spectra.  Employing 
a minimization algorithm, we find the linear combination of individual 
spectra, which, when suitably normalized and velocity-broadened, best fits the 
spectral regions in the object of interest that are free of emission lines.

The large number of galaxies in our survey presents us with an extensive 
collection of absorption-line spectra spanning a wide range of velocity 
dispersion, metallicity, and stellar population.  Since our survey is 
observationally unbiased with respect to the presence or absence of emission 
lines, fairly uniform signal-to-noise ratios (S/N) were achieved for all 
galaxies.  In total, there are 65 galaxies free of visible emission lines; an 
additional 14 have only very weak emission lines in the red spectrum and 
almost none in the blue spectrum.  The library of 79 templates (Table 1) was 
ranked according to velocity dispersion and approximate metallicity, the 
latter judged by the average equivalent widths of three strong \ion{Fe}{1} 
features in the blue spectrum (\S\ 4.2).  The diversity of the template 
library can be seen in the sample spectra shown in Figure 1.  

The $\chi^2$-minimization algorithm was adapted from a code originally 
developed to study the line-of-sight velocity distributions of early-type 
galaxies (see Rix \& White 1992 for a full description).  First, we corrected 
all the spectra, both the emission-line objects and the templates, for 
reddening due to the Galaxy (Burstein \& Heiles 1982, 1984) using the 
extinction law of Cardelli, Clayton, \& Mathis (1989).  This step removes 
gross continuum shape mismatches between the template and object spectra due 
to extrinsic dust.  Given a list of input template spectra and an initial 
guess of the velocity dispersion, the program solves for the systemic 
velocity, the line-broadening function (assumed to be a Gaussian), the relative 
contributions of the various templates, and the overall continuum shape.  In 
general, a good solution for the absorption-line spectrum does not guarantee 
an equally good match for the continuum shape, as the latter can be affected 
by dust intrinsic to the galaxies, by residual bandpass calibration errors, 
and by the non-uniqueness of the continuum shape as predicted by the strength 
of the stellar absorption lines.  The best-fitting model is then subtracted 
from the original spectrum, yielding a pure emission-line spectrum.  We 
typically chose from our master list one to several candidates whose velocity 
dispersions, metallicity, and range in stellar populations bracket the 
probable values in the object of interest.  We often iterated this process 
several times using different combinations of input templates.  Each iteration 
was applied separately to the blue and the red spectra, since the two are not 
contiguous. We selected the final, best model by inspecting the flatness of 
the residual spectrum in regions not containing strong emission lines.

Figure 2 illustrates this process for NGC 3596, whose nucleus is undergoing 
star formation, and for the Seyfert 2 nucleus of NGC 7743.  In the case of 
NGC 3596, the model consists of a combination of the spectrum of NGC 205, 
a dE5 galaxy dominated by A-type stars, and NGC 4339, an elliptical with a 
K-giant spectrum.  Note that in the original observed spectrum ({\it top}), 
H$\gamma$, \oiii\ \lamb\lamb 4959, 5007, and \oi\ \lamb 6300 are hardly 
visible, whereas they can be easily measured after starlight subtraction 
({\it bottom}).  The intensities of both H\bet\ and H\al\ were modified 
substantially, and the ratio of the doublet \sii\ \lamb\lamb 6716, 6731 lines 
changed.  The effective template for NGC 7743 also makes use of NGC 205 and 
NGC 4339, but in this case the intermediate-age population contributes 
substantially less; moreover, the fit improved after including a 
contribution from NGC 628, an Sc galaxy with a nucleus dominated by A and 
F stars.  Additional examples of starlight subtraction in Figures 3 and 4 
depict the range of emission-line intensities encountered in practice.  The 
weakest emission lines we can measure have equivalent widths of $\sim$0.25 
\AA, and we take this to be our detection limit (3 $\sigma$).

The philosophy behind our empirical method is simple and close in spirit to 
that of a similar technique often employed in the past (e.g., Costero \& 
Osterbrock 1977; Koski 1978; Shuder \& Osterbrock 1981; Peimbert \& 
Torres-Peimbert 1981; Rose \& Searle 1982; Filippenko \& Halpern 1984; 
Filippenko 1985; Filippenko \& Sargent 1988; Ho, Filippenko, \& Sargent 
1993).  The templates used in 
these previous studies, however, were generally derived either from the 
spectrum of a single, different galaxy with weak or undetectable emission 
lines or from the spectrum of an off-nuclear position in the same galaxy.  
Both cases, however, can suffer from significant shortcomings.  In the first 
instance, the absorption-line galaxy chosen as the template may not accurately 
reflect the stellar content and velocity dispersion of the object in 
question.  Most studies generally invest only limited observing time to the 
acquisition of template spectra, and the suitability of the choice of 
templates is not always stated explicitly.  In the second case, the 
off-nuclear spectrum may not be completely free of line emission, and, if 
there are radial gradients in the stellar population, the off-nuclear light 
may be a poor representation of the nuclear position.  Our implementation of 
the template-subtraction technique largely overcomes these difficulties. 
Instead of using a single (and often subjectively determined) spectrum as
the template, we use an objective algorithm to find the best combination of 
spectra to create an ``effective'' template.  The advantages of this 
modification are obvious.  Use of a large basis of input spectra ensures a 
closer match to the true underlying stellar population, while automation of 
the process minimizes subjective judgement and allows efficient processing of 
the large database.

Our method of starlight removal can be futher compared with those which 
generate the best-fitting model for the stellar spectrum from population 
synthesis (e.g., Keel 1983b).  Although physically more rigorous, the 
appropriateness of the latter approach may be questionable.  The available 
input stellar libraries are drawn mainly from stars with solar metallicity, 
while galaxy nuclei, especially those in early-type systems, tend to be 
metal-rich environments (Spinrad \& Taylor 1971).  Indeed, for many of his 
objects, Keel was unable to measure even relatively prominent emission lines 
such as H\bet.  In an effort to circumvent some of the problems encountered 
in traditional synthesis techniques, Bica and co-workers (Bica \& 
Alloin 1986; Bica 1988; Storchi-Bergmann, Bica, \& Pastoriza 1990) have 
compiled a spectral library of star clusters having a range of metallicities. 
 
Finally, a common practice involves fitting the Balmer absorption and emission
series simultaneously by assuming that all the absorption lines have the
same equivalent width (e.g., McCall, Rybski, \& Shields 1985; Dinerstein
\& Shields 1986; Liu \& Kennicutt 1995; Veilleux \etal 1995).  Although
H$\delta$, H$\gamma$, and H$\beta$ have roughly comparable equivalent widths
under a variety of conditions, H\al\ appears to show deviations
(D\'\i az 1988).  Moreover, the profiles of the Balmer absorption series are
pressure broadened, especially for a young to intermediate-age population;
failure to account for this in the line-fitting process, as seems to have
been the case in the above-mentioned studies, will severely compromise the
outcome of the line decomposition.  Moreover, in fitting only the Balmer 
lines, one implicitly assumes that stellar absorption does not affect the other
emission lines in the optical window.  This, however, is unjustified for old 
stellar systems: stellar features affect the majority of emission lines 
to some degree (e.g., Fig. 3 of Ho \etal 1993), and the effect can be severe 
if the emission lines are weak.
 
Column 13 of Table 2 lists the template(s) used for each of the 418 
emission-line galaxies in the survey.  Nonstellar, featureless emission 
dominates the continuum in several luminous Seyfert galaxies (e.g., NGC 1275, 
NGC 3227, NGC 4151), and starlight subtraction need not be applied for these.  
Likewise, the continuum is essentially featureless for a few objects with 
intense star-forming regions (e.g., IC 10, NGC 784, NGC 1569).  For these 
objects, we fitted the continuum with a low-order spline function and 
subtracted it prior to line measurement.

\subsection{Measurement of Emission Lines}

After starlight subtraction, the intensity, full width at half maximum (FWHM), 
and central wavelength were determined for all emission lines detected.  The 
prominent lines include H$\gamma$, H\bet, \oiii\ \lamb\lamb 4959, 5007, \oi\ 
\lamb\lamb 6300, 6363, \nii\ \lamb\lamb 6548, 6583, H\al, and \sii\ \lamb\lamb 
6716, 6731.  In many cases, the lines can be measured straightforwardly 
by fitting a single Gaussian to each line, with the following exceptions.  
(1) Profile fitting is unreliable for very noisy data, for lines that are 
affected by large residuals from poor starlight subtraction, or for lines with 
exceptionally complex structure, such as those showing boxy profiles or 
multiple velocity components.  In these cases, we resorted to summing the 
line intensity between two endpoints interactively specified.  (2) The 
strength of H\bet\ emission in a few objects is either unavailable (e.g., the 
corresponding pixels were damaged by cosmic rays) or otherwise very unreliable 
because of severe template mismatch.  Since our spectral classification of the 
nuclei (\S\ 3) depends on H\bet, we simply fixed its strength in these cases
based on that of H\al\ by adopting a theoretical ratio of the two lines, with 
the implicit assumption that the reddening is negligible.  (3) Approximately 
10\% of the sample exhibits broad H\al\ emission and requires more involved 
line decomposition; this subset of objects is discussed in detail by Ho \etal 
(1997e).  (4) A significant fraction of the galaxies, especially those with 
active nuclei, have lines with obviously asymmetric shapes and other 
deviations from a single Gaussian profile.  If warranted by the data quality, 
these cases were fitted with multiple Gaussian components to better 
characterize the parameters of the line profile.

We placed upper limits on the intensities of undetected, but diagnostically 
important, emission lines by simulating measurements of the noise level near 
the expected positions of the lines.  The widths of the measurement 
``windows'' were assumed to be comparable to the FWHMs of the lines which 
are detected in the same object.

Table 2 (cols. 2--8) summarizes the observed (not corrected for reddening) 
intensities of the main emission lines relative to the intensity of the narrow 
component of H\al; we chose the normalization relative to H\al\ instead of 
H\bet\ because the former is usually better determined. If the observing 
conditions were deemed to be approximately photometric (see Paper II), we list 
the H\al\ flux (col. 9) and the associated luminosity (col. 10) using the
distances adopted in Table 10; nonphotometric measurements are given as 
lower limits.  Column 11 lists the equivalent width of the narrow H\al\ 
emission line, obtained by dividing the H\al\ intensity by the continuum level 
at 6563 \AA.  
 
In some spectra, particularly those with sufficiently high S/N, one or more of 
the following weaker lines were also measurable, depending on the object: 
\oiii\ \lamb 4363, \hei\ \lamb 4471, \heii\ \lamb 4686, \siii\ \lamb 6312, 
\fex\ \lamb 6374, and \hei\ \lamb 6678.   The intensities of these lines, 
where detected, are summarized in Table 3.

Previous studies of the profiles of narrow emission lines (e.g., Heckman \etal 
1981; Feldman \etal 1982; Whittle 1985) generally use \oiii\ \lamb 5007 as the 
fiducial line, since it is normally the brightest line in the optical spectrum 
and is usually unaffected by blending with other lines.  In our survey, 
however, we will use \nii\ \lamb 6583 instead of \oiii\ \lamb 5007 as the 
fiducial line for line-width measurements, since \nii\ lies in the red 
spectrum, and the S/N and spectral resolution of the red spectra are superior 
to those of the blue spectra.   Despite being partly blended with H\al\ in 
some instances, \nii\ \lamb 6583 is normally the strongest line in the 
red spectrum next to H\al, and, unlike H\al, it is not strongly affected 
by potential uncertainties introduced by imperfect starlight subtraction.  
The FWHM values of \nii\ are given in column 12 of Table 2.  The FWHM is 
the simplest, albeit not necessarily the most reliable (e.g., Whittle 1985), 
indicator of the line width.  Nevertheless, we present the FWHM measurements 
here as a first-order parameterization of the emission-line kinematics; 
a more extensive analysis of the line profiles is deferred to a later 
paper.  The tabulated line widths were corrected for instrumental 
resolution by subtracting the instrumental width in quadrature from the 
observed width: FWHM$^2$ = FWHM$^2_{\rm obs}\,-$ FWHM$^2_{\rm inst}$.  
FWHM$_{\rm inst}$ is obtained by fitting a Gaussian profile to a strong, 
unblended emission line near 6583 \AA\ in the comparison lamp spectrum taken 
closest in time to the object of interest.\footnote{The median FWHM$_{\rm inst}$
for the entire survey is found to be 2.4 \AA, somewhat smaller than the 
nominal value of 2.5 \AA.  For completeness, we note that the median 
FWHM$_{\rm inst}$ of the blue spectra near \lamb\ = 5000 \AA\ is 3.8 \AA, 
again slightly better than the nominal resolution of 4 \AA.}  The lines in 
the comparison lamp spectra are very nearly Gaussian in shape.  To minimize 
the possibility of corrupting the final FWHM values from systematic errors in 
the determination of the instrumental profile, we only apply the correction if 
FWHM$_{\rm obs}$ exceeds FWHM$_{\rm inst}$ by at least 50\%.  Widths smaller 
than this are reported as upper limits.  

\subsection{Reddening Correction}

We adopt the Cardelli \etal (1989) extinction curve to correct the line 
intensities for reddening due to Galactic dust; the reddening values we 
assume, taken from Burstein \& Heiles (1982, 1984), are listed in column 3 of 
Table 11.  Next, we use the corrected H\al/H\bet\ intensity ratio (Table 4; 
col. 2) to calculate the internal reddening [$E(B-V)_{int}$; col. 3] due to 
dust which is either intrinsically associated with the line-emitting gas or 
lying in its foreground, but belonging to the host galaxy.  The Cardelli et 
al. extinction curve was again assumed.  For objects whose spectra resemble 
\hii\ regions, we take the intrinsic H\al/H\bet\ ratio to be 2.86, appropriate 
for case B recombination in photoionized nebulae with $n_e$ = 10$^2$ \cc\ and 
$T_e$ = 10$^4$ K (Brocklehurst 1971), whereas for AGNs we adopt an intrinsic 
ratio of 3.1 (Halpern \& Steiner 1983; Gaskell \& Ferland 1984).  Note that in 
some objects the observed Balmer decrement is significantly less than the 
theoretical value.  These obviously discrepant values, which are usually 
either lower limits or otherwise have a low quality rating, most likely can be 
attributed to inaccurate starlight subtraction, especially for H\bet.  We 
assign an internal extinction of zero for these objects.

Table 4 (cols. 4 and 6 through 9) lists five important diagnostic line 
ratios.  They have been corrected for Galactic and internal reddening, 
although in all cases the correction is nearly negligible because of the close 
wavelength separation of the lines used in the ratios.

\subsection{Electron Densities}

The electron density, $n_e$, can be estimated from the ratio $R$(\sii) $\equiv$ 
\sii\ \lamb 6716/\sii\ \lamb 6731.  $R$(\sii) serves as an effective 
densitometer (e.g., Osterbrock 1989) so long as $n_e$ does not greatly exceed 
the critical density of \sii\ ($\sim$3\e{3} \cc), above which the lines become 
collisionally de-excited.  We used the NEBULAR package implemented within 
IRAF\footnote{IRAF is distributed by the National Optical Astronomy
Observatories, which are operated by the Association of Universities for
Research in Astronomy, Inc., under cooperative agreement with the National
Science Foundation.} to compute $n_e$ (Table 4; col. 5) from $R$(\sii) (col. 
4), assuming $T_e$ = 10$^4$ K.  The software, described by Shaw \& 
Dufour (1995), assumes that the ground-state electron configuration of 
common ions such as S$^+$ can be approximated by a five-level atom.  The 
atomic data for S$^+$ are those published by Cai \& Pradhan (1993).

\subsection{Uncertainties and Error Estimates}

It is generally difficult to evaluate the uncertainty associated with the
starlight-subtraction process, which in almost all cases dominates the 
systematic errors in the line measurements.  Hence, formal errors are not 
assigned to the entries in Tables 2--4.  Instead, we employ less quantitative, 
but nevertheless informative, quality ratings (blank, ``b,'' and ``c'') to 
denote the probable reliability of the quoted values.  Throughout the tables, 
entries not followed by a flag (blank rating) are deemed to be of the best 
quality.  We estimate that, on average, the highest accuracy achieved for the 
line intensities is on the order of $\pm$10\%--30\%, depending on the line.  A 
quality rating of ``b'' signifies probable errors of $\pm$30\%--50\%, while a 
rating of ``c'' pertains to probable errors as large as $\pm$100\%.  Note that 
these estimates do not apply to absolute fluxes, which are discussed below. 
The assignment of the ratings was based on our experience with data of this 
type, on repeated measurements of objects having multiple observations, and on 
extensive experimentation with a range of template models.

Being relatively bright in the red region and not strongly affected 
by stellar absorption, the intensities of \nii\ and \sii\ can be determined to 
an accuracy of $\sim$10\% in the best cases.  \oi\ \lamb 6300, athough not 
crucially dependent on starlight subtraction, is generally much weaker, and 
it is difficult to achieve errors less than 20\%--30\%.  Although \oiii\ \lamb 
5007 is usually quite bright and not strongly affected by absorption, the blue 
spectra of our survey unfortunately have lower sensitivity than the red 
spectra; nevertheless, accuracies of 10\%--20\% can be attained for this 
line.  The Balmer series potentially suffers from the most serious uncertainty, 
whose magnitude becomes progressively larger from H\al\ to H$\gamma$.  Because 
of the wavelength coincidence between the emission and absorption components, 
and because the strength of the absorption component depends sensitively 
on the age of the stellar population, the intensities of the hydrogen emission 
lines can vary quite dramatically depending on the choice of template model.  
Thus, template mismatch can introduce rather large uncertainties in the 
strengths of the hydrogen emission lines.  On the other hand, we believe that 
we can ascertain the suitability of the chosen template model.  Since the 
hydrogen absorption lines usually have pressure-broadened, Voigt profiles, 
their wings are much broader than those of the sharper emission component, 
thus making it possible to judge the appropriateness of the template model: 
if the template undersubtracts the absorption, a shallow ``bowl'' is evident 
beneath the emission line; conversely, if oversubtraction occurs, a broad 
``hump'' would be noticeable.  Given that the equivalent widths of the Balmer 
absorption lines rarely exceed 2--3 \AA\ (see \S\ 4.2 and Table 9), it is 
reasonable to assume that emission lines with an equivalent width  
$\sim$10 \AA\ should have an error of at most 10\% for H\al\ and perhaps 20\% 
for H\bet.  

Although we provide emission-line fluxes and luminosities (Table 2) and 
continuum magnitudes (\S\ 4.1; Table 8) for about 80\% of the survey objects, 
these values do not have high accuracy for at least two reasons.  First, the 
observations were taken through a narrow slit of fixed width (2\asec).  
The narrow-line emission in many objects is known to be both spatially 
extended and inhomogeneous.  The amount of light included for each object 
obviously depends on the specific spatial distribution of the line emission 
and of the stellar continuum, on the accuracy of centering the object on 
the slit and subsequent telescope guiding, and on the seeing conditions.  
Moreover, because of the fixed slit size, the amount of light admitted into 
the aperture will depend on the distance of the source.  Second, we did not 
attempt to perform rigorous spectrophotometry; the sky conditions were merely 
monitored by eye periodically through the night.  To make a rough estimate of 
the accuracy of the fluxes, we compared all available measurements for objects 
which were observed using the same slit width and in more than one epoch, under 
purportedly ``photometric'' conditions.  Some of the (unpublished) data were 
collected with the Hale 5~m telescope and the double spectrograph, but at lower 
resolution, and others were obtained from Lick Observatory using the Shane 
3~m telescope and the Kast spectrograph (Miller \& Stone 1993), also at lower 
resolution.  The comparison, which could be made for about 30 galaxies, 
showed that the fluxes are generally consistent to within $\pm$50\% and never 
deviated by more than a factor of 2.  Therefore, we take this to be the 
range of internal uncertainty associated with each individual measurement, 
although the uncertainty in our subsequent analysis should be much less because 
we only consider the ensemble properties of the sample.  

The aperture effect, although expected to be present at some level, evidently 
is not very severe.  The distribution of H\al\ emission-line fluxes shows no 
obvious trend with source distance (Fig. 5).  The only discernible pattern,
due to the Malmquist bias, is the lack of points in the lower left corner of 
the diagram, implying a deficit of nearby (\lax 20 Mpc) high-luminosity 
sources.  Nevertheless, for the other reasons mentioned above, we still 
strongly caution the reader against using any individual measurement for 
quantitative applications.

Finally, we consider the reliability of the line-width measurements.  The 
merits of using \nii\ \lamb 6583 as the representative narrow emission line 
for kinematic measurements were discussed in \S\ 2.2.  The errors 
associated with the actual measurement of the FWHM should in general be 
\lax 10\% of the line width.  What is potentially more important are 
possible systematic biases on the line widths introduced during the 
observations.  Under good seeing conditions in which the width of 
the seeing disk is smaller than the width of the slit, and for two sources 
having identical intrinsic line widths, the observed line width of an 
unresolved source will be systematically narrower than that of an extended 
source, provided that the guiding errors do not exceed the slit width.  
This effect can therefore introduce a systematic bias in the measured line 
widths depending on the structure of the nuclear regions.  However, we 
anticipate that this source of error will be unimportant in our survey 
because the combination of guiding errors and seeing was comparable to the 
slit width.  We confirmed our hypothesis by examining the distribution of 
the seeing conditions (FWHM) recorded in the observation logs (Fig. 6); there 
are no systematic differences among the various classes of nuclei (\S\ 3), 
with the mean and median seeing FWHMs being essentially identical for the four 
classes.  More importantly, within each class, there are no discernible 
differences in the average FWHM(\nii) for the objects observed under ``good'' 
seeing compared to those observed under ``poor'' seeing, where we have taken 
the dividing line to be FWHM = 1\farcs5.  The only possible exception is the 
Seyfert group, where the average FWHM(\nii) = 274 and 241 \kms\ for seeing 
FWHM $<$ 1\farcs5 and $\geq$ 1\farcs5, respectively.  This trend, however, is 
opposite to that predicted, and, given the small sample size, is probably 
statistically insignificant.  

\section{Spectral Classification}

The optical spectral region contains several emission lines whose intensity 
ratios can be used to discriminate different sources of ionization.  
Baldwin, Phillips, \& Terlevich (1981) suggested several two-dimensional, 
line-intensity ratio diagrams that have since become widely-used diagnostic 
tools to classify emission-line objects.  In these diagrams, nebulae
photoionized by hot, young stars (\hii\ regions) can be distinguished from 
those photoionized by a harder radiation field, such as that of an AGN.
The Str\"omgren sphere of a classical, ionization-bounded \hii\ region makes a
fairly sharp transition between the ionized interior and the surrounding 
neutral medium.  Consequently, the partially-ionized transition zone, from 
which most of the low-ionization forbidden lines (such as \oi, \sii, and 
to some extent \nii) are emitted, is quite limited in size.  By contrast, 
an AGN-like ionizing continuum has sufficient high-energy photons to sustain 
an extensive partially-ionized zone in clouds optically thick to the Lyman 
continuum, thereby enhancing the strengths of the low-ionization lines 
relative to those seen in \hii\ regions.  A suitable choice of an excitation 
indicator, such as \oii\ \lamb 3727/\oiii\ \lamb 5007 or \oiii\ 
\lamb 5007/H\bet, further delineates the excitation sequence in \hii\ regions 
and divides AGNs into a high-excitation class  (Seyferts) and a low-excitation 
class [low-ionization nuclear emission-line regions (LINERs); Heckman 1980b].

The separation between the two principal ionization sources (young stars 
vs. AGN) and between the two AGN excitation classes (Seyferts vs. LINERs) 
obviously does not have sharp, rigorously-defined boundaries.  In practice, 
however, one must impose somewhat arbitrary, albeit empirically motivated, 
criteria to establish an internally consistent system of classification.
In this study, we use the diagnostic diagrams recommended by Veilleux \& 
Osterbrock (1987), since they employ line-intensity ratios that are relatively 
insensitive to reddening (\oiii\ \lamb 5007/H\bet, \oi\ \lamb 6300/H\al, 
\nii\ \lamb 6583/H\al, and \sii\ \lamb\lamb 6716, 6731/H\al) and that are 
contained in the spectral coverage of our survey.  For concreteness, we adopt 
the classification criteria outlined in Table 5; these are very similar to 
those used by Veilleux \& Osterbrock.  In addition to \hii\ nuclei, Seyferts, 
and LINERs, we also recognize a group of ``transition objects,'' noted by 
Ho \etal (1993) as emission-line nuclei having \oi\ strengths intermediate 
between those of \hii\ nuclei and LINERs.  Ho \etal (1993; see also Ho 1996) 
proposed that transition objects can be most naturally explained as ``normal'' 
LINERs whose integrated spectra are diluted or contaminated by 
neighboring \hii\ regions.  In this interpretation, transition objects should 
be considered part of the LINER class and should be accounted in AGN 
statistics (if all LINERs are truly AGNs), although an alternative explanation 
for these objects based solely on stellar photoionization has also been 
suggested (Filippenko \& Terlevich 1992; Shields 1992).  

As discussed in Ho \etal (1997e), broad H\al\ emission can be discerned in 
a significant number of objects in our survey.  Although the typical 
luminosity and contrast of the broad component are much smaller than those 
of ``classical'' type 1 Seyferts, the widths of the H\al\ line (FWHM $\approx$ 
1000--4000 \kms), especially relative to the forbidden lines, signify the 
presence of a broad-line region.  Although the existence of such 
``intermediate-type'' nuclei has been recognized for a long time (Osterbrock 
1977, 1981; Shuder 1980), it is normally thought that weak, broad H\al\ 
emission is widely found only in Seyfert nuclei.  The Palomar 
survey has demonstrated conclusively that broad H\al\ exists also in LINERs, 
and perhaps even in some transition objects.  Thus, analogous to the 
conventional nomenclature of Seyfert nuclei, we propose that the ``type 1'' 
and ``type 2'' designations, along with the intermediate types (1.2, 1.5, 1.8, 
and 1.9; Osterbrock 1981), be extended to include LINERs and transition 
objects.  We use the criteria of Whittle (1992a) to quantitatively assign 
numerical values for the type index.

We wish to stress that our classification system is defined by a set of 
{\it spectroscopic} criteria that depends entirely on the line-intensity 
ratios of several prominent, narrow, optical emission lines, rather than by a 
combination of spectroscopic, photometric, and morphological characteristics, 
as has been customary in many older studies.  The requirement that a Seyfert 
nucleus be bright and semistellar in appearance (e.g., de Vaucouleurs \& de 
Vaucouleurs 1968), for instance, in addition to being subjective, obviously 
cannot be extended to nuclei of very low luminosities, whose contrast relative 
to the galaxy bulge becomes the imperceptibly low.  Likewise, as pointed out 
by Phillips, Charles, \& Baldwin (1983), the traditionally accepted definition 
based on the {\it widths} of the emission lines (Weedman 1970, 1977; Balzano 
\& Weedman 1981; Shuder \& Osterbrock 1981) should be abandoned.  The 
widths of the emission lines in active nuclei correlate strongly with 
luminosity (Phillips \etal 1983; Whittle 1985, 1992b; Ho 1996; Ho \etal 
1997d), and imposing an arbitrary line-width cutoff selects objects by their 
luminosities.  Finally, one should resist the practice of automatically 
labeling objects as ``Seyfert 1s'' simply from their having emission lines 
(e.g., H\al) arising from a broad-line region (e.g., Paper I).  As mentioned 
above, broad lines {\it also} exist in objects with lower ionization levels, 
and these, according to their narrow-line spectra, would be classified as 
either LINERs or transition objects.

The last column of Table 4 lists the adopted classification for each of the 
418 emission-line nuclei, while Figures 7{\it a}--7{\it c} show the 
locations of the objects on the three Veilleux \& Osterbrock (1987) diagnostic 
diagrams.  Some representative spectra of each class, offset from each other 
for clarity, are plotted in Figures 8--12.  We detected emission-line nuclei 
in 86\% of the survey galaxies.  In total, there are 206 \hii\ nuclei, 94 
LINERs, 65 transition objects, and 52 Seyferts.  These make up, respectively, 
49\%, 23\%, 16\%, and 12\% of the emission-line objects and 42\%, 19\%, 13\%, 
11\% of the entire galaxy sample (Ho \etal 1997a discusses the overall 
statistics of the emission-line nuclei in greater detail).  One object 
(NGC 1003) could not be classified because of the large uncertainties in 
the line ratios.

It should be emphasized that the classification process is not always 
unambiguous, for at least three reasons.  First, the three conditions 
involving the low-ionization lines (\oi, \nii, and \sii) do not always hold 
simultaneously.  This reflects the empirical nature of the diagnostic 
diagrams as well as the possibility of any one line ratio to be enhanced 
or depressed with respect to the other two as a result of, for instance, 
selective abundance variations.   Second, substantial measurement uncertainty 
may be associated with any given line-intensity ratio.  Finally, one or more 
of the line ratios is sometimes only available as an upper limit.
Thus, each object must be evaluated individually, taking all of these 
factors into consideration, before a classification can be assigned to it. 
In view of the potential selective enhancement of nitrogen in galactic nuclei 
(e.g., Storchi-Bergmann \& Pastoriza 1989, 1990), less weight is given to the 
\nii/H\al\ ratio than to either \oi/H\al\ or \sii/H\al.  If a reliable 
measurement of \oi\ is available, \oi/H\al\ is given the largest weight, 
since, of the three low-ionization lines, it is the most sensitive to the 
shape of the ionizing spectrum.  Where more than one classification may be 
consistent with the data, both are given, with the more likely one, and the 
one finally adopted, listed first.  Some degree of subjectivity must 
unavoidably enter into the classification process, although we have tried 
throughout to be as consistent and unbiased as possible.

We should point out that the definition of LINERs adopted here differs from 
that originally proposed by Heckman (1980b), which is based exclusively on the 
relative strengths of \oi, \oii, and \oiii.  Nevertheless, inspection of the 
full optical spectra of LINERs (e.g., Ho \etal 1993) reveals that in practice 
the two sets of classification criteria generally identify the same objects.  
This merely reflects that \oiii/H\bet\ inversely correlates with \oii/\oiii\ 
for conditions of low excitation (see Fig. 2 in Baldwin \etal 1981).  Figure 
13 shows the distribution of \oi/\oiii\ for all objects classified as LINERs 
in our survey according to the definition adopted here and for which this line 
ratio is securely known.  Clearly, almost all of the objects have \oi/\oiii\ 
\gax 0.33, one of Heckman's criteria.

In general, our classifications should be much better determined than those 
currently available in the literature, and we recommend the classification 
given here for overlapping objects.  Previous studies of this kind (e.g., 
Stauffer 1982a, b; Keel 1983a, b; Phillips \etal 1986; V\'eron-Cetty \& 
V\'eron 1986; V\'eron \& V\'eron-Cetty 1986) often relied exclusively on 
\nii/H\al\ for classification: any nucleus having \nii/H\al\ larger than some 
limit, usually taken to be $\sim$0.6, was automatically assumed to be 
``active.''  As discussed above, basing the classification on just one line 
ratio can often be misleading, even if \nii/H\al\ statistically correlates 
well with \oi/H\al\ and \sii/H\al\ (Fig. 14).  The results are all the more 
unreliable if starlight subtraction has not been treated carefully or 
consistently, as was the case in some of the older surveys; undersubtraction 
of the stellar absorption lines, or failure to remove them altogether, will 
result in substantial overestimation of \nii/H\al.  Moreover, some of the 
above-mentioned studies did not have sufficient wavelength coverage to include 
H\bet\ and \oiii, thereby leaving the excitation of the objects unconstrained.

In her analysis of a sample of relatively bright Seyfert 2 nuclei and LINERs, 
Storchi-Bergmann (1991) showed that the \nii/H\al\ ratio depends 
systematically on aperture size or source distance --- larger apertures 
include more extranuclear emission from \hii\ regions, thereby diluting the 
nuclear spectrum and lowering the observed \nii/H\al\ ratio.  If this effect 
is severe, it can of course change the spectral classification from that of an 
AGN to an \hii\ nucleus, and the AGN fraction in a survey such as ours will be 
underestimated. Since our observations were acquired with a fixed slit size, 
we know that this distance-dependent effect must be present at some 
level, and we need to evaluate to what extent our classifications might be 
influenced by it.  Figure 15 plots the distribution of \nii/H\al\ values as 
a function of distance; all the emission-line objects are shown together as 
one group, as each of the four spectral classes separately, and with the 
three AGN classes summed into one.  Within the limited range of distances in 
our sample, the \nii/H\al\ ratio exhibits no pronounced variation with 
distance.  Variations in effective aperture size do not adversely affect our 
spectral classifications in the mean, although, of course, small errors may 
still affect individual objects.  

\section{Stellar Parameters}

Our starlight-subtraction procedure yields, as a by-product, the best-fitting 
continuum and absorption-line model for each galaxy processed in this manner.  
These model template spectra, together with the original set of templates, 
contain useful information pertaining to the stellar component of the nuclei.  
Because of the high S/N of our spectra, even relatively weak stellar 
absorption lines can be measured accurately.  Unfortunately, the limited 
wavelength coverage of our observing setup excludes several key stellar 
features (e.g., the blue coverage just misses the 4000 \AA\ break to the left, 
and \ion{Mg}{1} $b$ \lamb 5175 and \ion{Fe}{1} \lamb 5270 to the right); 
nevertheless, it does include a number of stellar indices that can be used 
as diagnostics of the stellar population.  These include the Balmer lines 
(H$\gamma$, H\bet, and H\al), the G band at 4300 \AA, \ion{Ca}{1} \lamb 4455 
and \ion{Ca}{1}+\ion{Fe}{1} \lamb 6495, \ion{Fe}{1} \lamb\lamb 4383, 4531, 
4668, and occasionally \heii\ \lamb 4686 emission from Wolf-Rayet stars.  The 
shape of the continuum additionally provides some constraints on the stellar 
content and the reddening.

Our survey is also a rich source of stellar velocity dispersions.  Because 
the full analysis of the velocity dispersions is rather lengthy and involved, 
we defer discussion of these measurements to a separate paper.

\subsection{Measurement of Continuum Indices}
The continuum flux density, $f_\lambda$, was determined at three points along 
the spectrum centered on the rest wavelengths of H$\gamma$ (4340 \AA), H\bet\ 
(4861 \AA), and H\al\ (6563 \AA).  In order to obtain the 
continuum values in a consistent, objective manner, the spectra were 
measured using an automated algorithm.  For each of the fiducial 
wavelengths, the mean intensities of two flanking, relatively line-free 
regions were determined, and the average of the two resulting values was
taken to be the continuum level at the center of the line.  Table 6 shows the 
wavelength ranges of the continuum regions adopted.  It is necessary to 
exclude the region containing the absorption line itself so as not to 
artificially depress the true continuum level.  Moreover, averaging over a 
sufficiently large wavelength window prevents the final value from being 
significantly affected by small-scale structure that may be present in 
the spectra.  The continuum measurements, expressed in magnitudes ($m_{44}$, 
$m_{49}$, $m_{66}$), are listed in Table 8 for all objects observed under 
photometric conditions, where $m$ = --2.5 log $f_\nu\,-48.6$, with $f_\nu$ in 
units of \flux\ \perhz\ (Oke \& Gunn 1983).

We further use the continuum magnitudes to define three ``color'' indices: 
$c(44-49)\,\equiv\,m_{44} - m_{49}$, $c(44-66)\,\equiv\,m_{44} - m_{66}$, and 
$c(49-66)\,\equiv\,m_{49} - m_{66}$.  Since the blue and red spectra for each 
galaxy were taken simultaneously, their relative calibration should be 
reliable and does not depend on sky conditions.  Hence, color indices are 
available for all objects.  

Lastly, an absolute magnitude, $M_{44}$, is given for the bluest band
using the distances assumed in Table 10 (see \S\ 5.2).

\subsection{Measurement of Equivalent Widths of Absorption Lines}
The definitions of the nine stellar absorption-line indices extracted from the 
spectra are given in Table 7 (see also Fig. 16 for a graphic representation 
of the definitions of the indices), and the results are tabulated in Table 9.  
In addition to the three individual iron indices, we also calculated an 
average iron index, $<$W(Fe)$>$ $\equiv\,{1\over 3}$[W(Fe4383) + W(Fe4531) 
+ W(Fe4668)], which should serve as a more robust indicator of the 
iron strength.

The measurement of equivalent widths, which requires determination of the 
flux of a spectral feature and its associated continuum level, can be 
affected by systematic errors depending on the S/N of the data, the strength 
of the feature, and the velocity dispersion.  If the feature of interest 
is weak or the data particularly noisy, for example, marking the endpoints 
of the feature and the appropriate continuum level can sometimes be ambiguous 
if done manually.  To ensure internal consistency in the data set, the 
equivalent widths were measured noninteractively.  Briefly, for each of 
the indices, the fluxes of three wavelength regions in the rest frame of 
the object were determined:  one region corresponds to the limits of the 
absorption feature of interest, while two regions on either side define two 
continuum levels (Table 7).  Interpolation of the two flanking continuum 
values gives the continuum level underneath the spectral feature, and the 
ratio of the integrated flux of the feature to the continuum flux yields the 
equivalent width.

\section{Compilation of Host Galaxy Properties}
Subsequent papers in this series will examine possible connections 
between the spectroscopic characteristics of the nuclei and the more global 
properties of the host galaxies.  In preparation for this analysis, we 
provide here a compilation of a number of host galaxy parameters (Tables 10 
and 11).  We follow closely the format and notation of Whittle (1992a), who has 
recently provided a comprehensive summary of the global properties of bright 
Seyfert galaxies, although we also list several quantities not given by 
Whittle.  We have relied heavily on the Third Reference Catalogue of Bright 
Galaxies (RC3; de Vaucouleurs \etal 1991) for much of the data because this 
catalog has paid considerable attention to synthesizing data from a variety 
of sources into a single system.  Since we will be primarily interested in 
comparative analysis, the most important consideration is to have at hand 
a consistent, homogeneous database.  A short description of each parameter 
tabulated is given below.

\subsection{Morphological Type}
We list for each galaxy the mean revised morphological type and the mean 
numerical Hubble type index (T), as given in the RC3.  The morphological 
classification of the RC3 parallels the system in the first two editions of the 
catalog, which itself is based on the principles outlined by de Vaucouleurs 
(1959, 1963).  In this system, T = --6 to --4 correspond to ellipticals (E), 
T = --3 to --1 to lenticulars (S0), T = 0 to 9 to spirals (S0/a--Sm), and 
T = 10 to Magellanic irregulars (Im).  T = 90 designates non-Magellanic 
irregulars or ``amorphous'' galaxies (I0; Sandage \& Brucato 1979), and T = 99 
imples a ``peculiar'' morphology that cannot be fitted into any of the above 
categories.  Among spirals, those that are strongly or weakly barred are 
designated by the family symbols ``B'' and ``AB,'' respectively, and nonbarred 
members by ``A''.  Note that the morphological types given in Paper II were 
taken from the catalog of nearby galaxies of Kraan-Korteweg (1986), which 
relies heavily on the RSA.  In future analyses, we will use the morphological 
types from the RC3 as listed in Table 10, instead of those from the RSA given 
in Paper II, largely because the RC3 system recognizes intermediate bar types 
(AB), but also because it is more current.

The distribution of Hubble types for the survey sample is summarized in 
Table 13 and in Figure 17.  It is apparent that the sample contains virtually 
every morphological type, and hence provides an excellent representation of the 
general galaxy population. Since the objects were drawn from a 
magnitude-limited catalog, however, our sample certainly is not immune to the 
usual underrepresentation of low surface brightness systems.  Ellipticals 
comprise 12\% of the sample, while lenticulars, spirals, irregulars, and 
peculiar systems make up 18\%, 65\%, 3\%, and 1.5\%, respectively.  Barred 
galaxies (AB and B) contribute 56\% to the disk systems (S0--Im) and 
59\% of the spiral galaxies in the sample, roughly comparable to the overall 
representation in the field galaxy population (50\%--60\%; Sellwood \& 
Wilkinson 1993).

\subsection{Distances}
The majority of the objects in our survey (90\%) overlap with the nearby 
galaxy catalog of Tully (1988), who derived distances for galaxies 
closer than 40 Mpc based on the Virgo infall model of Tully \& Shaya (1984).  
This model assumes an infall velocity of 300 \kms\ for the Local Group, 
$H_0$ = 75 \kms\ Mpc$^{-1}$, and a Virgo distance of 16.8 Mpc.  Although some 
of the galaxy distances have been slightly modified in the recent literature, 
we chose to adopt Tully's distances for convenience and for internal 
consistency.  For galaxies farther than 40 Mpc, we obtained distances 
simply from their systemic velocities and $H_o$ = 75 \kms\ Mpc$^{-1}$.  The 
distribution of distances in our survey (Fig. 18) has a median value of 
17.9 Mpc, a mean of 22.8 Mpc, and a standard deviation of 14.9 Mpc.  The 
aperture used in the spectral extractions (2\asec\ $\times$ 4\asec) 
corresponds to a projected linear size of 180 pc $\times$ 360 pc for the 
median distance.  The radial velocities of the galaxy nuclei, which are 
available from our data, will be presented elsewhere along with additional 
kinematic data.

%XX
%Technically, I think I need to use the velocity wrt the Local Group to 
% calculate distances from H_o.  But for distances > 40 Mpc or v > 3000 km/s, 
% not correcting for this only makes a difference of about 3-4\% on the 
% final velocities, which is within the error of the velocities anyways. 
% Rather than waste a lot of time trying to do this correction and then 
% changing all the distance dependent quantities, I will simply neglect it.

\subsection{Radial Velocities and Rotational Velocities}
The heliocentric radial velocities, $v_{\rm h}$, are listed in column 4 
of Table 10; these were taken from the RC3, after converting the values 
given therein in the frame of the Galactic Standard of Rest to the frame 
of the Sun.  They represent the weighted average of the best available 
redshifts measured through 21-cm neutral hydrogen and optical observations. 
The \hi\ redshifts for the Palomar sample have typical uncertainties of 
$\sim$6 \kms, while the optical measurements have somewhat larger ones, 
generally $\sim$30 \kms.  A histogram of $v_{\rm h}$ is shown in Figure 19.

The amplitude of the rotation curve of a spiral galaxy provides a 
measure of its gravitational potential, and hence its total mass.
Previous studies of Seyfert galaxies successfully employed the rotational 
amplitude to assess the role of the bulge in influencing the nuclear
activity (e.g., Whittle 1992b), and we plan to do the same for the objects 
in this survey.  Although our spectra were acquired in long-slit mode, the 
spectrograph was usually not rotated to the photometric major axis of each 
galaxy, but rather to the parallactic angle to minimize light losses due to 
atmospheric dispersion (see Paper II).  Hence, only some of the resulting 
rotation curves (to be published elsewhere) can be used to derive reliable 
rotation amplitudes.  Fortunately, it has been demonstrated (e.g., Rubin, 
Ford, \& Thonnard 1978; Whittle 1992a) that the maximum rotational velocity 
inferred from the width of the integrated \hi\ emission profile of spiral 
galaxies provides an excellent substitute for the rotational amplitude obtained 
through direct analysis of an optical rotation curve.  Whittle (1992a) showed 
that W20, the width of the \hi\ profile measured at 20\% of the peak height, 
correlates extremely well with $\Delta$V$_{\rm rot}$, the amplitude of the 
optical rotation curve, and that the correlation apparently holds 
for both normal and 
active galaxies.  Accordingly, we have taken the available W20 measurements 
from the RC3 for our survey objects (largely based on the \hi\ database of 
Bottinelli \etal 1990), and these are tabulated as  $\Delta$V$_{\rm HI}$ in 
Table 10.  Since the rotational velocities are seen in projection along the 
line of sight, the true, inclination-corrected rotational amplitudes
$\Delta$V$^c_{\rm rot}$ = $\Delta$V$_{\rm HI}/$sin $i$ are also listed for 
all the disk systems, where we have used the inclinations given in 
Table 11 (see \S\ 5.8).  To ensure that the deprojection correction factor is 
not excessively large, inclinations less than 30\deg\ were set to 30\deg.

\subsection{\hi\ Content}
It will be of interest to search for possible connections between the presence 
and type of nuclear activity and the overall gaseous content of the host 
galaxies, as some relationship may be found based on the expectation that 
the availability of interstellar matter should affect the fueling rate of 
the nuclear regions.  Since the RC3 conveniently lists \hi\ fluxes for the 
majority of the non-elliptical galaxies in our survey, we summarize the 
\hi\ masses in Table 10, normalized to the ``corrected'' $B$-band luminosity 
(see \S\ 5.6) of each object, $M_{\rm HI}/L^0_B$, where in converting 
magnitudes to luminosities a solar absolute magnitude of $M_B$ = 5.52 was
used.  The calculation of the \hi\ mass from the observed \hi\ fluxes is 
explained in the RC3, and, in the present compilation, we simply adopted the 
fluxes given in the RC3, which account for self-absorption of the 21-cm line.
Note that the quantity $M_{\rm HI}/L^0_B$ is independent of distance.

\subsection{Galaxy Environment}
That tidal perturbations from neighboring galaxies can trigger the onset of 
nuclear activity as well as enhance its strength has been documented in 
numerous studies (e.g., Shlosman 1994 and references therein).  For each 
galaxy, we attempt to gauge the gravitational influence of its nearest 
neighbors using two estimators.  First, we make use of the local galaxy 
density, $\rho_{\rm gal}$, defined by Tully (1988) as the density of all 
galaxies 
brighter than $M_B$ = --16 mag in the vicinity of the object of interest.  A 
second, more straightforward approach that has proven to be useful in the 
past (e.g., Stauffer 1982b) is to measure the separation between a given 
galaxy and that of its nearest, sizable neighbor.  A reasonable strategy might 
be to locate a neighbor meeting a relative magnitude criterion ($\Delta m$), 
as well as a relative velocity criterion ($\Delta v$).  The former selects 
neighbors falling within a given luminosity ratio, which loosely translates 
into a mass ratio if the luminosity is not abnormally enhanced or depressed, 
whereas the latter increases the probability that the pair is actually 
physically associated.  We chose $\Delta m\,\leq 1.5$ mag (factor of 4 in 
luminosity) and $\Delta v\,=\,\pm$500 \kms, and we performed an automated 
search  using the NASA/IPAC Extragalactic Database (NED) to measure 
$\theta_p$, the projected angular separation between the primary and the 
secondary galaxy, measured in units of the isophotal angular diameter of 
the primary, $D_{25}$ (Table 11, col. 11).  The search radius was limited 
to 300\amin, the maximum value allowed by the NED search routine; beyond this 
limit the galaxy can be safely considered to be isolated.  

Finally, we consider the environment of each galaxy in a larger context, namely 
its group and cluster membership.  Garcia (1993) recently analyzed 
the group affiliation for a large sample of nearby galaxies, whose selection
criteria encompasses those of the Palomar survey.  More than 66\% (323/486) 
of our objects can be assigned to specific groups, and they are identified in 
Table 10 using the group designation of Garcia.  Those objects belonging to the 
Virgo Cluster or its southern extension (Tully 1987) are additionally noted.

\subsection{Magnitudes and Colors}
The total (asymptotic) $B$-band magnitudes, as observed ($B_T$) and after 
correcting for extinction and redshift ($B_T^0$), were taken from the RC3 
and are listed in Table 11.  The two dominant terms in the magnitude 
correction are due to internal extinction ($A_i$), which depends on 
inclination and is assumed to be significant only for spirals (see RC3), and 
extinction from the Galaxy, $A_g$, which was estimated principally from the 
\hi\ maps of Burstein \& Heiles (1982, 1984).  Because of the proximity of 
our sample, the redshift ($K$) correction in almost all cases amounts to less 
than 0.01 mag.  The absolute magnitudes, $M_{B_T}^0$ = $B_T^0\,-$ 5 log 
($d$/10) [$d$ in pc], are derived from the apparent magnitudes and assumed 
distances.  The distributions of $B_T$ and $M_{B_T}^0$ are shown in Figures 20 
and 21, respectively.  
%Note that a small number of objects (XX, or XX\%) have apparent 
%magnitudes that formally fall outside of the magnitude limit of our survey 
%($B_T\,\leq$ 12.5 mag).

We also present in Table 11 two broad-band optical colors taken from the RC3, 
$(U-B)_T^0$ and $(B-V)_T^0$, integrated over the galaxy and corrected for 
the effects of extinction and redshift as described above.  These quantities 
will be of use for examining global trends in the star-forming properties 
of the host galaxies.

\subsection{Bulge Luminosities}
The mass of the galactic bulge or spheroidal component most likely exerts the 
strongest impact on nuclear activity, as the central engine undoubtedly 
is coupled to the depth of the nuclear gravitational potential.  To first 
order, such an association is already apparent in that AGNs preferentially 
reside in early-type hosts (Heckman 1980a; Keel 1983b; Terlevich, Melnick, 
\& Moles 1987; Ho \etal 1997a), in that the kinematics of the line-emitting 
gas reflect principally the velocities of the bulge stars (Nelson \& Whittle 
1996), and in the preliminary indication that more massive black holes appear 
to live in more luminous galaxies (Kormendy \& Richstone 1995).  While an 
explicit bulge-disk decomposition would certainly be desirable, at the moment 
detailed surface photometry is unavailable for our sample as a whole.  We 
follow Whittle (1992a) and use an average, type-dependent relation to estimate 
the bulge luminosity from the total galaxy luminosity.  Simien \& de 
Vaucouleurs (1986) give the following empirical relation to estimate the 
contribution of the disk to the total luminosity of a disk galaxy with 
--3 $\leq$ T $\leq$ 7: $\Delta m_{\rm bul}$ = 0.324(T + 5) -- 0.054(T + 5)$^2$ 
+ 0.0047(T + 5)$^3$.  This correction is zero for ellipticals (T = --6 to --4), 
which by definition have no disks, and it equals 1.02 mag for T = +1 (Sa) and 
2.54 mag for T = +5 (Sc).  In this study, we assume that the empirical 
relation can be extended to T = 9; for T $>$ 9, $\Delta m_{\rm bul}$ is 
undefined.  The bulge absolute magnitudes, $M_B$(bul) = $M_{B_T}^0$ + 
$\Delta m_{\rm bul}$, are listed in Table 11 and shown in histogram form in 
Figure 22.

\subsection{Inclinations and Isophotal Diameters}

It is instructive to examine the distribution of galaxy inclination angles 
to assess possible selection effects.  Figure 23 shows the distribution of 
cosine $i$ for all sample disk galaxies for which the RC3 lists an isophotal 
axial ratio, $R_{25}$ = log($a/b$), where the major and minor axis diameters, 
$a$ and $b$, are measured at a $B$ surface brightness level of 25 mag 
arcsec$^{-2}$.  The inclination then follows from sin$^2 i\,=\, 
[1 - (b/a)^2]/0.96$ (Hubble 1926).  The distribution of cosine $i$ should be 
flat for an unbiased sample with random orientations.  Our sample evidently 
is not completely free from inclination bias, as judged by the increasing 
deficit of objects for $i$ \gax 70\deg; the departure from a uniform 
distribution is highly significant according to a Kolmogorov-Smirnov test 
(Press \etal 1986).  This behavior, however, is to be expected in a 
magnitude-limited sample, since internal absorption tends to shift edge-on 
disk systems above the limiting magnitude (e.g., Burstein, Haynes, \& Faber 
1991; Maiolino \& Rieke 1995).

We also extracted from the RC3 the isophotal major axis diameters 
(measured at $\mu_B$ = 25 mag arcsec$^{-2}$), both as observed ($D_{25}$) 
and corrected to face-on orientation ($D^0_{25}$).  This parameter will be 
used as a measure of the linear scale of the systems.

\subsection{Far-infrared Properties}

Finally, we collected the best available far-IR
measurements of the host galaxies made by the {\it Infrared Astronomical 
Satellite (IRAS)}.  These data will be used to analyze the global 
star-formation properties of the galaxies.  The total flux densities 
($S_{12}$, $S_{25}$, $S_{60}$, and $S_{100}$) and the associated errors 
($\sigma$) of the four bands at 12, 25, 60, and 100 \micron, all in units of 
Janskys, are given in columns 2 through 9 of Table 12.  The parameter 
$FIR$ (col. 10), defined as 1.26\e{-14}(2.58$S_{60}\,+\,S_{100}$) W m$^{-2}$, 
approximates well the total flux between 42.5 and 122.5 \micron\ 
(Rice \etal 1988; Helou \etal 1988).  For convenience, we also tabulated the 
the ratios of the flux densities of the four bands, or the far-IR colors: 
$S_{12}/S_{25}$, $S_{25}/S_{60}$, and $S_{60}/S_{100}$ (cols. 11--13).

Many catalogs of {\it IRAS} data exist in the literature.  We have strived 
to adopt the largest, most homogeneous, and most recent compilations 
available.  Flux densities for galaxies with diameters larger than 8\amin\ 
were taken from Rice \etal (1988), and the remaining ones were chosen, in 
decreasing order of preference, from Helou \etal (1988), Soifer \etal (1989), 
Knapp \etal (1989), Rush, Malkan, \& Spinoglio (1993), and Moshir \etal 
(1990), the latter as given in NED.  Additional studies (col. 14) were 
consulted for a few objects not listed in the above primary sources.  (The 
published version of the Rush et al. catalog did not list uncertainties for 
the flux densities.  B. Rush kindly provided the original data files 
from which we extracted the uncertainty measurements.)

\section{Summary}

In an effort to search for and to study the physical properties of 
low-luminosity AGNs, we have completed an optical spectroscopic survey of 
the nuclear regions of a large sample of nearby galaxies.  The survey is useful 
for many other purposes in addition to the original objectives.  This paper 
describes in detail our procedure of measuring various spectroscopic 
parameters to characterize the emission lines, the absorption lines, 
and the stellar continuum.  From analysis of the emission-line intensity 
ratios, spectral classifications are presented for a total of 418 
emission-line nuclei.  We discuss in depth our procedure for starlight 
subtraction, as this is a crucial step for accurate measurement of the 
relatively weak emission lines in these nuclei.  Finally, we provide a 
summary of global galaxy properties compiled from various published catalogs.  
This information will be used extensively in other papers in this series, 
where the scientific analysis will be presented.

\acknowledgments

The research of L.~C.~H. is currently funded by a postdoctoral fellowship
from the Harvard-Smithsonian Center for Astrophysics.  Financial support for 
this work was provided by NSF grants AST-8957063 and AST-9221365, as well as 
by NASA grants AR-5291-93A and AR-5792-94A from the Space Telescope Science 
Institute (operated by AURA, Inc., under NASA contract NAS5-26555).  We thank 
Hans-Walter Rix for generously providing the code used for the starlight 
subtraction and Chien Peng for developing the software used for the 
line-profile fitting. Brian Rush gave valuable advice concerning {\it IRAS} 
data and sent us an electronic version of the extended 12 \micron\ galaxy 
catalog.  Anne Manuelle Garcia furnished an electronic file of 
her catalog of galaxy groups.  Chris McKee and Hy Spinrad gave a critical 
reading of an earlier draft of the manuscript, and helpful input during the 
final revision of the paper was provided by Bill Keel, Kim McLeod, 
Roberto Maiolino, and an anonymous referee. Throughout the course of this 
work, L.~C.~H. has benefited from discourse with Aaron Barth and Tom Matheson, 
and well as from spirited consumption of caffeine with Marianne Vestergaard.
We made extensive use of the NASA/IPAC Extragalactic Database (NED)
which is operated by the Jet Propulsion Laboratory, California Institute
of Technology, under contract with the National Aeronautics and Space
Administration.

\clearpage
\appendix
\section{Notes on Individual Objects}

Here we record special difficulties, peculiarities, and other 
miscellaneous notes of interest encountered during the starlight subtraction 
or in subsequent steps of the analysis.

{\it  NGC  278.} --- Starlight subtraction in the blue spectrum is uncertain. 
The H\bet\ and H$\gamma$ absorption lines could not be completely removed. 

{\it  NGC  428.} --- Spectral classification is very uncertain.

{\it  NGC  474.} --- Spectral classification is very uncertain.

{\it  NGC  488.} --- The emission lines have double peaks separated by $\sim$200 
\kms.

{\it  NGC  521.} --- \oiii\ \lamb 5007 is very poorly measured.

{\it  NGC  1073.} --- Starlight subtraction in the blue spectrum is uncertain. 
The H\bet\ and H$\gamma$ absorption lines could not be completely removed. 

{\it  NGC  1560.} --- \oi\ \lamb 6300 is corrupted by emission from the night 
sky.

{\it  NGC  2366.} --- The spectrum is corrupted by contamination from a 
superposed M star, and the object is discarded from the sample.

{\it  NGC  2832.} --- Starlight subtraction very difficult. Spectral 
classification highly uncertain.

{\it  NGC  3034.} --- The continuum shape of the red spectrum was
improperly calibrated, although this does not affect the measurement of the 
emission lines.

{\it  NGC  3077.} --- The H\bet\ and H$\gamma$ absorption lines could not be 
completely removed during starlight subtraction.

{\it  NGC  3379.} --- Starlight subtraction very difficult. We could not 
obtain a reasonable value for H\bet\ emission and resorted to constraining it 
with H\al.

{\it  NGC  3608.} --- Starlight subtraction very difficult. We could not 
obtain a reasonable value for H\bet\ emission and resorted to constraining 
it with H\al.

{\it  NGC  3646.} --- The emission lines have a double-peaked profile.
 
{\it  NGC  3665.} --- The emission lines have a double-peaked profile.
 
{\it  NGC  3884.} --- \sii\ \lamb 6731 corrupted.
 
{\it  NGC  3941.} --- H\bet\ strength uncertain by factor $\sim$2.  
Classification could change from Seyfert 2 to LINER 2.

{\it  NGC  4145.} --- The spectrum is partly contaminated by an M star.  
Although we could not remove the stellar component, we were able to 
estimate the strengths of several emission lines.

{\it  NGC  4220.} --- H\bet\ may be higher than quoted by a factor of $\sim$2, 
but this will not change the spectral classification.

{\it  NGC  4278.} --- The \oiii\ \lamb 4363 line quoted in Table 3 may be 
partly contaminated by residuals from H$\gamma$.

{\it  NGC  4281.} --- The emission lines are very noisy and have complicated 
line profiles.  We summed several of the key emission lines by manual 
integration.

{\it  NGC  4350.} --- Starlight subtraction very difficult. We could not 
obtain a reasonable value for H\bet\ emission and resorted to constraining it 
with H\al.  The \sii\ lines were summed as one complex and then divided into 
a ratio of 1.5 to 1 for \sii\ \lamb 6716 and \sii\ \lamb 6731, respectively.

{\it  NGC  4414.} --- H\al\ emission corrupted; estimated based on poorly 
constrained strength of H\bet.

{\it  NGC  4472.} --- Starlight subtraction very difficult. H\bet\ emission 
strength obtained from H\al, which itself is very uncertain.  The \sii\ lines 
were summed as one complex and then divided into a ratio of 1.5 to 1 for 
\sii\ \lamb 6716 and \sii\ \lamb 6731, respectively.

{\it  NGC  4490.} --- The stellar population is very young, and the Balmer 
absorption lines are most likely undercorrected.  The H\al/H\bet\ ratio 
should be regarded as an upper limit.

{\it  NGC  4579.} --- H\bet\ was damaged by a cosmic ray hit; its intensity 
was scaled from that of H\al\ emission.
 
{\it  NGC  5322.} --- Starlight subtraction was very difficult and the choice 
of template ambiguous.  The final line intensities represent the average of 
5 independent trials.  The two lines of \sii\ could not be measured 
individually; we summed the blend and divided the intensity equally between the 
two lines, as the emission complex looks relatively flat-topped, indicating 
that the two lines probably contribute in equal proportion.

{\it  NGC  5631.} --- The Balmer emission lines, especially H\bet, are somewhat 
uncertain, but in this case the spectral classification should not be affected.

{\it  NGC  5656.} --- The Balmer emission lines are poorly determined.  
\sii\ \lamb 6731 is partly corrupted and therefore not reliable.

{\it  NGC  5858.} --- The spectrum is very noisy, and the line profiles are 
quite complex.  The line strengths in the red were approximated by eye, 
and the ratio of the \sii\ lines is not meaningful.

{\it  NGC  5982.} --- Starlight subtraction extremely difficult; line 
intensities and spectral classification uncertain.

{\it  NGC  6482.} --- The blue spectrum shown in Figure 94 of Paper II, 
which was obtained at Lick Observatory, cannot be that of NGC 6482. 
The velocity dispersion of the absorption lines in the blue spectrum is 
inconsistent with that of the red spectrum, and in any event appears to be 
too small for a giant elliptical galaxy like NGC 6482.  The blue spectrum 
was therefore discarded, and we have no information on the excitation 
of the emission lines.  Judging by the strength of \nii\ and \sii\ compared 
to H\al\ and by the relative weakness of \oi, we estimate that the nucleus 
is probably either a transition object or a Seyfert 2.

{\it  NGC  6501.} --- Starlight subtraction extremely difficult; line 
intensities and spectral classification uncertain.

{\it  NGC  6702.} --- Starlight subtraction extremely difficult.

%REFERENCES
\clearpage

\centerline{\bf{References}}
\medskip

\refindent
Baldwin, J.~A., Phillips, M.~M., \& Terlevich, R. 1981, \pasp, 93, 5

\refindent
Balzano, V.~A., \& Weedman, D.~W. 1981, \apj, 243, 756

\refindent
Bica, E. 1988, \aa, 195, 76

\refindent
Bica, E., \& Alloin, D. 1986, \aa, 162, 21

\refindent
Bottinelli, L., Gouguenheim, L., Fouqu\'e, P., \& Paturel, G. 1990, \aas, 
82, 391

\refindent
Brocklehurst, M. 1971, \mnras, 153, 471

%\refindent
%Bruzual A., G. 1983, \apj, 273, 105

\refindent
Burstein, D., Haynes, M.~P., \& Faber, S.~M. 1991, \nat, 353, 515

\refindent
Burstein, D., \& Heiles, C. 1982, \aj, 87, 1165

\refindent
Burstein, D., \& Heiles, C. 1984, \apjs, 54, 33

\refindent
Bushouse, H.~A., Lamb, S.~A., \& Werner, M.~W. 1988, \apj, 335, 74

\refindent
Cai, W., \& Pradhan, A.~K. 1993, \apjs, 88, 329

\refindent
Cardelli, J.~A., Clayton, G.~C., \& Mathis, J.~S. 1989, \apj, 345, 245

\refindent
Costero, R., \& Osterbrock, D.~E. 1977, \apj, 211, 675

\refindent
de Vaucouleurs, G. 1959, Handbuch der Physik, 53, 275

\refindent
de Vaucouleurs, G. 1963, \apjs, 8, 31

\refindent
de Vaucouleurs, G., \& de Vaucouleurs, A. 1968, \aj, 73, 858

\refindent
de Vaucouleurs, G., de Vaucouleurs, A., Corwin, H.~G., Jr., Buta, R.~J., 
Paturel, G., \& Fouqu\'e, R. 1991, Third Reference Catalogue of Bright 
Galaxies (New York: Springer) (RC3)

\refindent
D\'\i az, A.~I. 1988, \mnras, 231, 57

\refindent
Dinerstein, H.~L., \& Shields, G.~A. 1986, \apj, 311, 45

\refindent
Feldman, F.~R., Weedman, D.~W., Balzano, V.~A., \& Ramsey, L.~W. 1982, \apj,
256, 427

\refindent
Filippenko, A.~V. 1985, \apj, 289, 475

\refindent
Filippenko, A.V., \& Halpern, J.~P. 1984, \apj, 285, 458

\refindent
Filippenko, A.~V., \& Sargent, W.~L.~W. 1985, \apjs, 57, 503 (Paper I)

\refindent
Filippenko, A.~V., \& Sargent, W.~L.~W. 1988, \apj, 324, 134

\refindent
Filippenko, A.~V., \& Terlevich, R. 1992, \apj, 397, L79

\refindent
Garcia, A.~M. 1993, \aas, 100, 47

\refindent
Gaskell, C.~M., \& Ferland, G.~J. 1984, \pasp, 96, 393

\refindent
Halpern, J.~P., \& Steiner, J.~E. 1983, \apj, 269, L37

\refindent
Heckman, T.~M. 1980a, \aa, 87, 142

\refindent
Heckman, T.~M. 1980b, \aa, 87, 152

\refindent
Heckman, T.~M., Miley, G.~K., van Breugel, W.~J.~M., \& Butcher, H.~R. 1981,
\apj, 247, 403

\refindent
Held, E.~V., \& Mould, J.~R. 1994, \aj, 107, 1307

\refindent
Helou, G., Khan, I.~R., Malek, L., \& Boehmer, L. 1988, \apjs, 68, 151

\refindent
Ho, L.~C. 1996, in The Physics of LINERs in View of Recent Observations, ed.
M. Eracleous, et al. (San Francisco: ASP), 103

\refindent
Ho, L.~C., Filippenko, A.~V., \& Sargent, W.~L.~W. 1993, \apj, 417, 63 

\refindent
Ho, L.~C., Filippenko, A.~V., \& Sargent, W.~L.~W. 1995, \apjs, 98, 477 
(Paper II)

\refindent
Ho, L.~C., Filippenko, A.~V., \& Sargent, W.~L.~W. 1997a, \apj, in press

\refindent
Ho, L.~C., Filippenko, A.~V., \& Sargent, W.~L.~W. 1997b, \apj, in press

\refindent
Ho, L.~C., Filippenko, A.~V., \& Sargent, W.~L.~W. 1997c, \apj, in press

\refindent
Ho, L.~C., Filippenko, A.~V., \& Sargent, W.~L.~W. 1997d, in preparation

\refindent
Ho, L.~C., Filippenko, A.~V., Sargent, W.~L.~W., \& Peng, C.~Y. 1997e, 
\apjs, in press

\refindent
Hubble, E. 1926, \apj, 64, 321

\refindent
Keel, W.~C. 1983a, \apjs, 52, 229

\refindent
Keel, W.~C. 1983b, \apj, 269, 466

\refindent
Knapp, G.~R., Guhathakurta, P., Kim, D.-W., \& Jura, M. 1989, \apjs, 70, 329

\refindent
Kormendy, J., \& Richstone, D.~O. 1995, \annrev, 33, 581

\refindent
Koski, A.~T. 1978, \apj, 223, 56

\refindent
Kraan-Korteweg, R.~C. 1986, \aas, 66, 255

\refindent
Liu, C.~T., \& Kennicutt, R.~C. 1995, \apjs, 100, 325

\refindent
Maiolino, R., \& Rieke, G.~H. 1995, \apj, 454, 95

\refindent
Mazzarella, J.~M., \& Balzano, V.~A. 1986, \apjs, 62, 751

\refindent
McCall, M.~L., Rybski, P.~M., \& Shields, G.~A. 1985, \apjs, 57, 1

\refindent
Melisse, J.~P.~M., \& Israel, F.~P. 1994, \aas, 103, 391

\refindent
Miller, J.~S., \& Stone, R.~P.~S. 1993, Lick Obs. Tech. Rep., No. 66

\refindent
Moshir, M., \etal 1990, IRAS Catalogs, The Faint Source Catalog, Version 2.0
(Pasadena, JPL)

\refindent
Nelson, C.~H., \& Whittle, M. 1996, \apj, 465, 96

\refindent
Oke, J.~B., \& Gunn, J.~E. 1983, \apj, 266, 713

\refindent
Osterbrock, D.~E. 1977, \apj, 215, 733

\refindent
Osterbrock, D.~E. 1981, \apj, 249, 462

\refindent
Osterbrock, D.~E. 1989, Astrophysics of Gaseous Nebulae and Active
Galactic Nuclei (Mill Valley: Univ. Science Books)

\refindent
Peimbert, M., \& Torres-Peimbert, S. 1981, \apj, 245, 845

\refindent
Phillips, M.~M., Charles, P.~A., \& Baldwin, J.~A. 1983, \apj, 266, 485

\refindent
Phillips, M.~M., Jenkins, C.~R., Dopita, M.~A., Sadler, E.~M., \&
Binette, L. 1986, \aj, 91, 1062

\refindent
Press, W.~H., Flannery, B.~P., Teukolsky, S.~A., \& Vetterling, W.~T. 1986,
Numerical Recipes (Cambridge: Cambridge Univ. Press)

%\refindent
%Pritchet, C.~J. 1977, \apjs, 35, 397

\refindent
Rice, W., Lonsdale, C.~J., Soifer, B.~T., Neugebauer, G., Kopan, E.~L.,
Lloyd, L.~A., de Jong, T., \& Habing, H.~J. 1988, \apjs, 68, 91

\refindent
Rix, H.-W., \& White, S.~D.~M. 1992, \mnras, 254, 389

\refindent
Roberts, M.~S., Hogg, D.~E., Bregman, J.~N., Forman, W.~R., \& Jones, C.
1991, \apjs, 75, 751

\refindent
Rodr\'\i guez-Espinosa, J.~M., Rudy, R.~J., \& Jones, B. 1987, \apj, 312, 555

\refindent
Rose, J.~A., \& Searle, L. 1982, \apj, 253, 556

\refindent
Rubin, V.~C., Ford, W.~K., Jr., \& Thonnard, N. 1978, \apj, 225, L107

\refindent
Rush, B., Malkan, M.~A., \& Spinoglio, L. 1993, \apjs, 89, 1

\refindent
Sandage, A.~R., \& Brucato, R. 1979, \aj, 84, 472

\refindent
Sandage, A.~R., \& Tammann, G.~A. 1981, A Revised Shapley-Ames Catalog of
Bright Galaxies (Washington, DC: Carnegie Institute of Washington) (RSA)

\refindent
Sanders, D.~B., Egami, E., Lipari, S., Mirabel, I.~F., \& Soifer, B.~T. 1995,
\aj, 110, 1993

%\refindent
%Sargent, W.~L.~W., \& Filippenko, A.~V. 1991, \aj, 102, 107

\refindent
Sellwood, J.~A., \& Wilkinson, A. 1993, Rep. Prog. Phys., 56, 173

\refindent
Shaw, R.~A., \& Dufour, R.~J. 1995, \pasp, 107, 896

\refindent
Shields, J. C. 1992, \apj, 399, L27

\refindent
Shlosman, I. 1994, ed., Mass Transfer Induced Activity in
Galaxies, (Cambridge: Cambridge Univ. Press)

\refindent
Shuder, J.~M. 1980, \apj, 240, 32

\refindent
Shuder, J.~M., \& Osterbrock, D.~E. 1981, \apj, 250, 55

\refindent
Simien, F., \& de Vaucouleurs, G. 1986, \apj, 302, 564

\refindent
Soifer, B.~T., Boehmer, L., Neugebauer, G., \& Sanders, D.~B. 1989, \aj, 98, 766

%\refindent
%Spinrad, H. 1962, \apj, 135, 715

\refindent
Spinrad, H., \& Taylor, B.~J. 1971, \apjs, 22, 445

\refindent
Stauffer, J.~R. 1982a, \apjs, 50, 517

\refindent
Stauffer, J.~R. 1982b, \apj, 262, 66

\refindent
Storchi-Bergmann, T. 1991, \mnras, 249, 404

\refindent
Storchi-Bergmann, T., Bica, E., \& Pastoriza, M.~G. 1990, \mnras, 245, 749

\refindent
Storchi-Bergmann, T., \& Pastoriza, M.~G. 1989, \apj, 347, 195

\refindent
Storchi-Bergmann, T., \& Pastoriza, M.~G. 1990, \pasp, 102, 1359

\refindent
Surace, J.~A., Mazzarella, J., Soifer, B.~T., \& Wehrle, A.~E. 1993, \aj, 105,
864

\refindent
Terlevich, R., Melnick, J., \& Moles, M. 1987, in Observational Evidence of
Activity in Galaxies, ed. E.~Ye. Khachikian, K.~J. Fricke, \& J. Melnick
(Dordrecht: Reidel), 499

\refindent
Tully, R.~B. 1987, \apj, 321, 280

\refindent
Tully, R.~B. 1988, Nearby Galaxies Catalog (Cambridge: Cambridge Univ. Press)

\refindent
Tully, R.~B., \& Shaya, E.~J. 1984, \apj, 281, 31

%\refindent
%Turnrose, B.~E. 1976, \apj, 210, 33

\refindent
van Driel, W., de Graauw, Th., de Jong, T., \& Wesselius, P.~R. 1993, \aas,
101, 207

\refindent
Veilleux, S., Kim, D.-C., Sanders, D.~B., Mazzarella, J.~M., \& Soifer, B.~T.
1995, \apjs, 98, 171

\refindent
Veilleux, S., \& Osterbrock, D.~E. 1987, \apjs, 63, 295

\refindent
V\'eron, P., \& V\'eron-Cetty, M.-P. 1986, \aa, 161, 145

\refindent
V\'eron-Cetty, M.-P., \& V\'eron, P. 1986, \aas, 66, 335

\refindent
Weedman, D.~W. 1970, \apj, 159, 405

\refindent
Weedman, D.~W. 1977, \annrev, 15, 69

\refindent
Whittle, M. 1985, \mnras, 213, 1

\refindent
Whittle, M. 1992a, \apjs, 79, 49 

\refindent
Whittle, M. 1992b, \apj, 387, 109

\refindent
Whittle, M. 1992c, \apj, 387, 121

\refindent
Worthey, G. 1992, Ph.D. thesis, Univ. of California, Santa Cruz

%FIGURE CAPTIONS
\clearpage

\centerline{\bf{Figure Captions}}
\medskip

Fig. 1. --- 
Sample absorption-line spectra used as templates for starlight 
subtraction. The two discontinuous wavelength regions represent the blue and 
red spectra of our observation setup.  Arbitrary constants were added to 
the scaled spectra for clarity

Fig. 2. --- 
Illustration of the method of starlight subtraction.  In each panel, the top 
plot shows the observed spectrum, the middle plot shows the best-fitting 
``template'' model (offset by a constant) used to match the stellar component, 
and the bottom plot represents the difference between the object spectrum and 
the model spectrum.  In the case of the \hii\ nucleus of NGC 3596 ({\it a}), the
model was constructed from NGC 205 and NGC 4339, while for the Seyfert 2
nucleus of NGC 7743 ({\it b}), the model was derived from a linear combination
of NGC 205, NGC 628, and NGC 4339.

Fig. 3. --- 
Additional examples of starlight subtraction.  ({\it a}) NGC 1052.
({\it b}) NGC 2768.

Fig. 4. --- 
Additional examples of starlight subtraction.  ({\it a}) NGC 3389.
({\it b}) NGC 4419.

Fig. 5. --- 
Flux of the narrow H\al\ emission line plotted against the 
galaxy distance.  There is no obvious correlation between the two quantities. 
Only data of quality ``b'' or better are shown.

Fig. 6. --- 
Distributions of FWHM for \nii\ \lamb 6583 as a function of the seeing 
conditions (FWHM of the seeing disk in seconds of arc).  Only FWHM(\nii) values 
of quality ``b'' or better are shown. The top panel combines measurements for 
all emission-line objects, while the bottom four panels plot each of the 
spectral classes individually.

Fig. 7. --- 
({\it a}) Diagnostic diagram plotting log \oiii\ \lamb 5007/H\bet\
versus log \nii\ \lamb 6583/H\al\ for \hii\ nuclei ({\it crosses}),
Seyfert nuclei ({\it squares}), LINERs ({\it solid circles}), and transition 
objects ({\it open circles}).  ({\it b}) Diagnostic diagram plotting log
\oiii\ \lamb 5007/H\bet\ versus log \sii\ \lamb\lamb 6716, 6731/H\al.
Symbols same as in ({\it a}).  ({\it c}) Diagnostic diagram plotting log
\oiii\ \lamb 5007/H\bet\ versus log \oi\ \lamb 6300/H\al.  Symbols same as in
({\it a}).

Fig. 8. --- 
Sample starlight-subtracted spectra of \hii\ nuclei.

Fig. 9. --- 
Sample starlight-subtracted spectra of LINERs.

Fig. 10. --- 
Sample starlight-subtracted spectra of transition objects.

Fig. 11. --- 
Sample starlight-subtracted spectra of Seyfert 2 nuclei.

Fig. 12. --- 
Sample starlight-subtracted spectra of Seyfert 1 nuclei.

Fig. 13. --- 
Distribution of \oi\ \lamb 6300/\oiii\ \lamb 5007 for the 64 LINERs in the 
Palomar survey with reliable measurements of both lines.  Line ratios 
having quality rating ``c'' and all upper limits were excluded.

Fig. 14. --- 
Correlations among the low-ionization optical forbidden lines.  
({\it a}) Log \nii\ \lamb 6583/H\al\ vs. log \oi\ \lamb 6300/H\al. 
({\it b}) Log \sii\ \lamb 6725/H\al\ vs. log \oi\ \lamb 6300/H\al\ 
(\sii\ \lamb 6725 $\equiv$ \sii\ \lamb\lamb 6716, 6731).  Data with quality 
``b'' or better are shown as {\it filled dots}, and those with quality 
``c'' are shown as {\it open dots}.

Fig. 15. --- 
Distribution of \nii\ \lamb 6583/H\al\ as a function of galaxy distance.
Only \nii/H\al\ values of quality ``b'' or better are shown. The top panel 
combines measurements for all emission-line objects, while the bottom five 
panels plot each of the spectral classes separately, as well as the three 
AGN classes combined.

Fig. 16. --- 
Definition of the stellar absorption-line indices for the ({\it a}) blue and 
({\it b}) red spectral regions.

Fig. 17. --- 
Distribution of morphological types for the Palomar survey.  The {\it top} 
panel shows all 486 galaxies in the sample, the {\it middle} panel the 252 
unbarred galaxies, and the {\it bottom} panel the 234 barred (AB+B) galaxies.
The bottom abscissa gives the Hubble types, and the top abscissa the
corresponding T indices, both binned slightly following the convention in 
Table 13.

Fig. 18. --- 
Distribution of distances.  The width of each bin corresponds to 5 Mpc.

Fig. 19. --- 
Distribution of heliocentric radial velocities.  The width of each bin 
corresponds to 500 \kms.

Fig. 20. --- 
Distribution of total apparent blue magnitudes ($B_T$).
The width of each bin corresponds to 0.5 mag.

Fig. 21. --- 
Distribution of absolute blue magnitudes for the entire galaxy ($M_{B_T}^0$), 
corrected for Galactic and internal extinction. The width of each bin 
corresponds to 0.5 mag.

Fig. 22. --- 
Distribution of absolute blue magnitudes of the bulge component [$M_B$(bul)]. 
The width of each bin corresponds to 0.5 mag.

Fig. 23. --- 
Distribution of the cosine of the galaxy inclination angles ($i$). 
The width of each bin corresponds to 0.1 in cos $i$.  

%FIGURES
%\clearpage

%TABLES
\clearpage
\centerline{\bf {NOTE CONCERNING THE TABLES}}
\bigskip

Please note that the following tables are not included in this preprint: 
Tables 2, 3, 4, 8, 9, 10, 11, and 12.  These tables are very large, 
containing nearly 40,000 data entries. The entries are being checked for 
accuracy and occasionally still being revised.  We will not distribute 
these tables prior to the actual publication of the paper.  
We apologize for any inconvenience.

%\clearpage
%\begin{figure}
%\plotone{tables/table1.ps}
%\end{figure}
%
%\clearpage 
%\begin{figure}
%\plotone{tables/table5.ps}
%\end{figure} 
% 
%\clearpage 
%\begin{figure}
%\plotone{tables/table6.ps}
%\end{figure} 
% 
%\clearpage 
%\begin{figure}
%\plotone{tables/table7.ps}
%\end{figure} 
%
%\clearpage 
%\begin{figure}
%\plotone{tables/table13.ps}
%\end{figure} 
 
\end{document}